\documentclass[aps,prd,floatfix,12pt]{revtex4}
\usepackage{epsfig}
\usepackage{bm}
\usepackage{dcolumn}
\usepackage{latexsym}

\begin{document}
   
\preprint{\rightline{ANL-HEP-PR-10-46}}
   
\title{Thermodynamics of lattice QCD with 3 flavours of colour-sextet quarks.}

\author{J.~B.~Kogut}
\affiliation{Department of Energy, Division of High Energy Physics, Washington,
DC 20585, USA}
 \author{\vspace{-0.2in}{\it and}}
\affiliation{Dept. of Physics -- TQHN, Univ. of Maryland, 82 Regents Dr.,
College Park, MD 20742, USA}
\author{D.~K.~Sinclair}
\affiliation{HEP Division, Argonne National Laboratory, 9700 South Cass Avenue,
Argonne, IL 60439, USA}

\begin{abstract}
We have been studying QCD with 2 flavours of colour-sextet quarks to distinguish
whether it is QCD-like or conformal. For comparison we are now studying QCD
with 3 flavours of colour-sextet quarks, which is believed to be conformal
in the chiral limit. Here we present the results of simulations of lattice QCD
with 3 colour-sextet quarks at finite temperatures on lattices of temporal
extent $N_t=4$ and $6$, with masses small enough to yield access to the chiral
limit. As for the 2-flavour case, we find well-separated deconfinement and
chiral-symmetry restoration transitions, both of which move to appreciably
weaker couplings as $N_t$ is increased from $4$ to $6$. If this theory is
conformal, we would expect there to be a bulk chiral transition at a fixed
coupling. For this reason we conclude that for $N_t=4$ and $6$, the chiral and
hence the deconfinement transitions are in the strong-coupling domain where
the theory is essentially quenched. The similarity between the behaviours of
the 2 and 3 flavour theories suggested that the $N_t=4$ and $6$ transitions
for the 2-flavour theory also lie in the strong-coupling domain. The phase 
structure of both theories is very similar.

\end{abstract}

\maketitle

\newpage

\section{Introduction}

With the LHC starting to probe the Higgs sector of the standard model, studies
of models of this sector are timely. We are particularly interested in 
extensions of the standard model with strongly-coupled (composite) Higgs
sectors. The most promising of these are Technicolor theories, QCD-like gauge
theories with massless fermions, whose pion-like excitations play the role of
the Higgs field, giving masses to the $W$ and $Z$
\cite{Weinberg:1979bn,Susskind:1978ms}.
It has been pointed out that phenomenological difficulties which plague
Technicolor theories and their extensions can largely be avoided if the
fermion content is such that the running coupling constant evolves very slowly
(``walks'') over a considerable range of renormalization scales. Such
non-perturbative behaviour is best studied by lattice gauge theory simulations.
\cite{Holdom:1981rm,Yamawaki:1985zg,Akiba:1985rr,Appelquist:1986an}

The most promising candidates for gauge theories which ``walk'', are those 
where the one-loop contribution to the Callan-Symanzik $\beta$ function
implies asymptotic freedom, while the two-loop contribution has the opposite
sign. If these two terms describe the physics, the $\beta$ function has a
second zero at non-zero coupling. Such a zero would be an infrared (IR) fixed 
point for massless quarks, implying that the theory is conformal. However, if a
chiral condensate forms before the would-be IR fixed point is reached, the
running coupling starts to increase again and the fixed point is avoided. The
theory is then QCD-like, but the presence of a nearby fixed point means that
there is a region where the coupling constant evolves very slowly and the 
theory walks.

We have been studying QCD with colour-sextet quarks. If the number of flavours
$N_f=2$ or $3$, the theory is asymptotically free and the two-loop term in
the $\beta$ function has the opposite sign from the one-loop term. $N_f=2$ is
a candidate walking theory, provided that it is QCD-like. Phenomenological
studies such as those of \cite{Sannino:2004qp,Dietrich:2005jn} and the
references these contain, indicate that further investigations of this
2-flavour theory as a model of walking technicolor, are warranted. Since
asymptotic freedom is lost at $N_f=3\frac{3}{10}$, two-loop perturbation theory 
predicts that $N_f=3$ has an IR fixed point at a small enough value of the 
coupling constant that one might trust perturbation theory. For this reason it 
is believed that the $N_f=3$ theory probably has an IR fixed point, and is thus
a conformal field theory. It is therefore useful to compare the behaviour of
the $N_f=3$ and $N_f=2$ theories to see whether they show qualitative
differences. We have performed extensive lattice simulations of the $N_f=2$
theory at finite temperatures to try and determine whether it is conformal or
QCD-like \cite{Kogut:2010cz,Kogut:2011ty}. 
These simulations are continuing \cite{Sinclair:2011ie}.
Work on the $N_f=2$ theory using different actions is being performed by other 
groups \cite{Shamir:2008pb,DeGrand:2008kx,DeGrand:2009hu,DeGrand:2010na,%
degrand_lattice11,Fodor:2008hm,Fodor:2011tw,kuti_lattice2011}.
However, a consensus as to whether the theory is conformal or QCD-like has yet
to be achieved. In this paper we present simulations of lattice QCD with 3
colour-sextet quarks using the same methods as for the 2-flavour case, for
comparison.
     We simulate lattice QCD with 3 flavours of staggered colour-sextet quarks
at finite temperature on lattices with $N_t=4$ and $6$, using the deconfinement
and chiral-symmetry restoration transitions to study the evolution of the
running coupling constant. As in the 2-flavour theory we find widely separated
deconfinement and chiral-symmetry restoration temperatures. Both transitions
move to appreciably weaker couplings as $N_t$ is increased from $4$ to $6$.
Except that the $N_f=3$ transitions are, as expected, at stronger couplings
than their $N_f=2$ counterparts, the two theories behave very similarly. Since
we believe the $N_f=3$ theory to be conformal, the weak coupling conformal
domain should be separated from the chirally broken region by a bulk chiral
transition. Thus the coupling at the chiral transition should be fixed. This
strongly suggests that the $N_t=4$ and $6$ transitions are in the 
strong-coupling domain where the fermions are bound into a chiral condensate
at distances of order the lattice spacing or less, so that the theory is
controlled by quenched dynamics. The evolution of the couplings at the
transitions between $N_t=4$ and $6$ occurs because these are
finite-temperature transitions of the effectively-quenched theory. It was this
observation that suggested to us that the $N_t=4$ and $6$ transitions of the
$N_f=2$ theory might also be in the strong-coupling domain. This was born out
when we determined the position of its $N_t=8$ chiral transition
\cite{Kogut:2011ty}.

     The $N_f=3$ theory shows a clear 3-state signal in the phase of the
Wilson Line(Polyakov Loop) just above the deconfinement transition. At even
larger values of $\beta=6/g^2$, the 2 states with complex Wilson Lines disorder
into a state with a negative Wilson line. This phase structure is very similar
to that observed for the $N_f=2$ theory.

     In section~2 we define our lattice action and discuss our simulation
methods. Section~3 describes our simulations and results for $N_t=4$, while
section~4 is devoted to our $N_t=6$ simulations. We present our discussion
and conclusions in section~5.

\section{Methodology}

This section gives a description of our action and simulation methods given
in an almost identical form in our earlier publications 
\cite{Kogut:2010cz,Kogut:2011ty}.

For the gauge fields we use the standard Wilson (plaquette) action:
\begin{equation}
S_g=\beta \sum_\Box \left[1-\frac{1}{3}{\rm Re}({\rm Tr}UUUU)\right].
\end{equation}
For the fermions we use the unimproved staggered-quark action:
\begin{equation}
S_f=\sum_{sites}\left[\sum_{f=1}^{N_f/4}\psi_f^\dagger[D\!\!\!\!/+m]\psi_f
\right],
\end{equation}
where $D\!\!\!\!/ = \sum_\mu \eta_\mu D_\mu$ with 
\begin{equation}
D_\mu \psi(x) = \frac{1}{2}[U^{(6)}_\mu(x)\psi(x+\hat{\mu})-
                            U^{(6)\dagger}_\mu(x-\hat{\mu})\psi(x-\hat{\mu})],
\end{equation}
where $U^{(6)}$ is the sextet representation of $U$, i.e. the symmetric part of
the tensor product $U \otimes U$. When $N_f$ is not a multiple of $4$ we 
use the fermion action:
\begin{equation}
S_f=\sum_{sites}\chi^\dagger\{[D\!\!\!\!/+m][-D\!\!\!\!/+m]\}^{N_f/8}\chi.
\end{equation}
The operator which is raised to a fractional power is positive definite and
we choose the real positive root. This yields a well-defined operator. We 
assume that this defines a sensible field theory in the zero lattice-spacing
limit, ignoring the rooting controversy. (See for example \cite{Sharpe:2006re}
for a review and guide to the literature on rooting.) 

We use the RHMC method for our simulations \cite{Clark:2006wp}, where the
required powers of the quadratic Dirac operator are replaced by diagonal
rational approximations, to the desired precision. By applying a global
Metropolis accept/reject step at the end of each trajectory, errors due to the
discretization of molecular-dynamics time are removed.

Finite temperature simulations are performed by using a lattice of finite
extent $N_t$ in lattice units in the Euclidean time direction, and of infinite
extent $N_s$ in the spatial direction. In practice this means we choose
$N_s \gg N_t$. The temperature $T=1/N_ta$, where $a$ is the lattice spacing.
(In our earlier equations we set $a=1$.) Since the deconfinement temperature 
$T_d$ and the chiral symmetry restoration temperature $T_\chi$ should not
depend on $a$, and since $a=1/N_tT$, measuring the coupling $g$ at $T_d$ or
$T_\chi$ as a function of $N_t$ gives $g(a)$ for a series of $a$ values which
approach zero as $N_t \rightarrow \infty$. If the ultraviolet behaviour of the
theory is governed by asymptotic freedom, $g(a)$ should approach zero as 
$a \rightarrow 0$, i.e. $N_t \rightarrow \infty$. However, for the 3-flavour
theory under consideration in this paper, we expect that the chiral transition
is a bulk transition. If so, the coupling at the chiral transition should be
a constant, independent of $N_t$. (Since the $\beta$ value at the deconfinement
transition ($\beta_d$) is expected to be less than that at the chiral
transition ($\beta_\chi$), it follows that $\beta_d$ will also approach a
finite value as $N_t \rightarrow \infty$.)

We determine the position of the deconfinement transition as that value of
$\beta$ where the magnitude of the triplet Wilson Line (Polyakov Loop) increases
rapidly from a very small value as $\beta$ increases. The chiral phase 
transition is at that value of $\beta$ beyond which the chiral condensate
$\langle\bar{\psi}\psi\rangle$ vanishes in the chiral limit. Because we are
forced to simulate at finite quark mass, this value is difficult to determine
directly. We therefore estimate the position of the chiral transition by
determining the position of the peak in the chiral susceptibility 
$\chi_{\bar{\psi}\psi}$ as a function of quark mass, and extrapolating to
zero quark mass. The chiral susceptibility is given by 
\begin{equation}
\chi_{\bar{\psi}\psi} = V\left[\langle(\bar{\psi}\psi)^2\rangle
                      -        \langle\bar{\psi}\psi\rangle^2\right]
\label{eqn:chi}
\end{equation}
where the $\langle\rangle$ indicates an average over the ensemble of gauge
configurations and $V$ is the space-time volume of the lattice. Since the
fermion functional integrals have already been performed at this stage, this
quantity is actually the disconnected part of the chiral susceptibility. Since
we use stochastic estimators for $\bar{\psi}\psi$, we obtain an unbiased
estimator for this quantity by using several independent estimates for each
configuration (5, in fact). Our estimate of $(\bar{\psi}\psi)^2$ is then given
by the average of the (10) estimates which are `off diagonal' in the noise.

\section{$N_t=4$ simulations}

We simulate lattice QCD with 3 flavours of sextet quarks on $12^3 \times 4$
lattices starting from an $N_f=2$ configuration with a positive Wilson Line
(Polyakov Loop) at $\beta=6/g^2=7.0$. Three different quark mass values 
$m=0.02$, $m=0.01$ and $m=0.005$ (in lattice units) are used to allow 
continuation to the chiral limit. For $m=0.02$ we choose a set of $\beta$s 
covering the range $5.0 \le \beta \le 7.0$, for $m=0.01$, 
$5.2 \le \beta \le 7.0$, while for $m=0.005$, $5.27 \le \beta \le 7.0$. This 
covers both the deconfinement and chiral-symmetry restoration transitions. For 
most values of $(\beta,m)$ we find 10,000 trajectories to be adequate. We 
increase this to 50,000 trajectories for values of $(\beta,m)$ close to the 
deconfinement transition.

\begin{figure}[hb]
\epsfxsize=6in
\epsffile{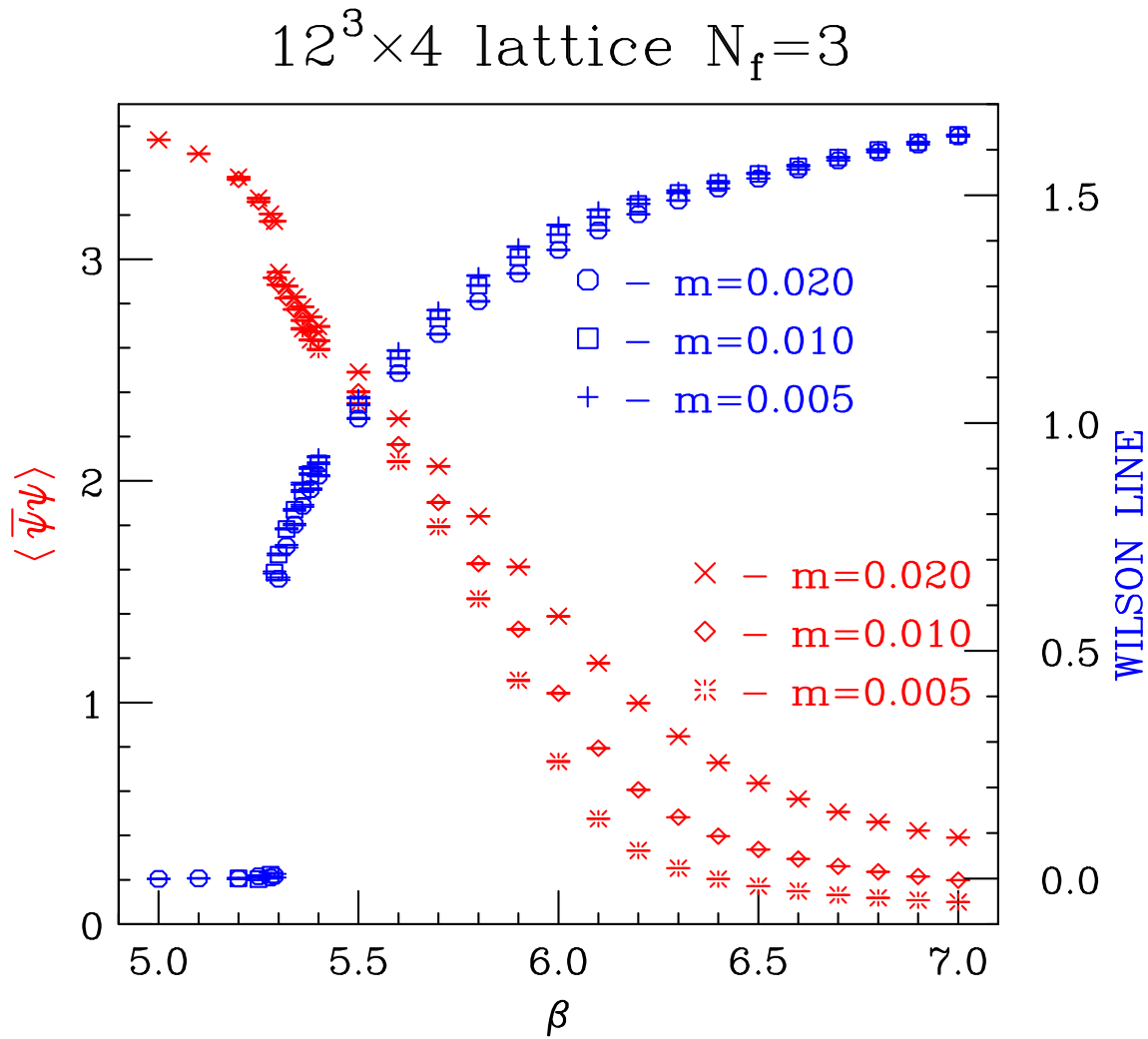}
\caption{Wilson Line (Polyakov Loop) and $\langle\bar{\psi}\psi\rangle$
as functions of $\beta$ on a $12^3 \times 4$ lattice.}
\label{fig:wil-psi_12x4}
\end{figure}

Figure~\ref{fig:wil-psi_12x4} shows the Wilson Lines and chiral condensates
$\langle\bar{\psi}\psi\rangle$ as functions of $\beta$ for each of the 3 mass
values used in our simulations. The qualitative features of these plots are
very clear. At low $\beta$ values, the Wilson Line is close to zero. At 
$\beta \approx 5.3$ or just below, the Wilson Line exhibits a (possibly 
discontinuous) jump to much larger values. This, we interpret to herald the
deconfinement transition. We note, however, that the chiral condensate, while
showing a small discontinuity, shows no sign of vanishing even in the chiral
limit, at this transition. Hence the deconfinement and chiral transitions are
not coincident, in contrast to what happens for fundamental quarks where all
the evidence favours chiral-symmetry restoration {\it at} the deconfinement 
transition. Pinpointing the chiral transition is more difficult, since this
requires extrapolating the chiral condensate to zero quark mass in the region
where it has strong dependence on the quark mass, to find where it vanishes.
A cursory examination of the figure would suggest that this occurs somewhere
around $\beta=6$. Hence the chiral and deconfinement transitions are far apart
with the chiral phase transition occurring at a much weaker coupling than the 
deconfinement transition.

\begin{figure}[htb]
\epsfxsize=3.2in
\epsffile{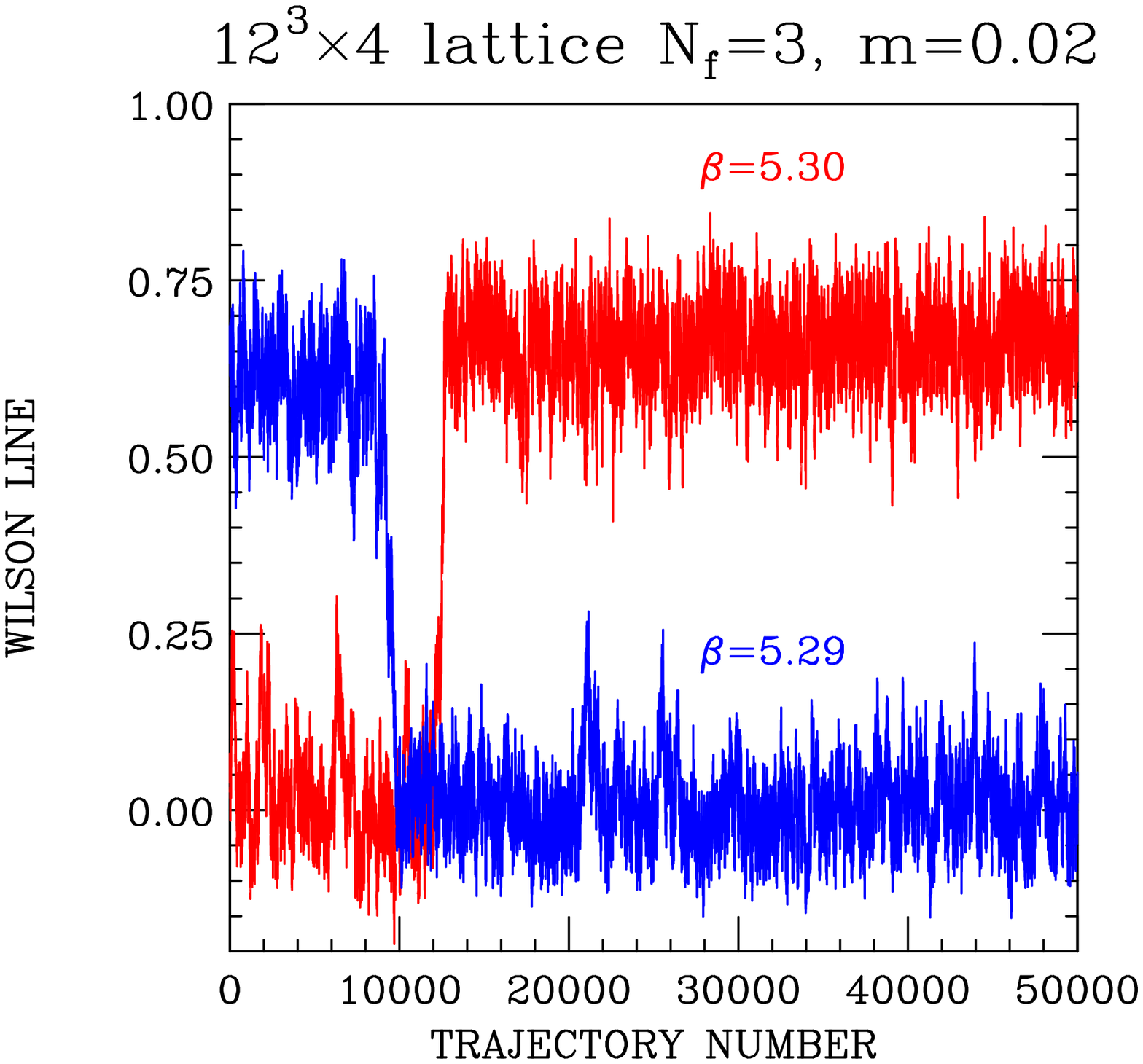}
\epsfxsize=3.2in
\epsffile{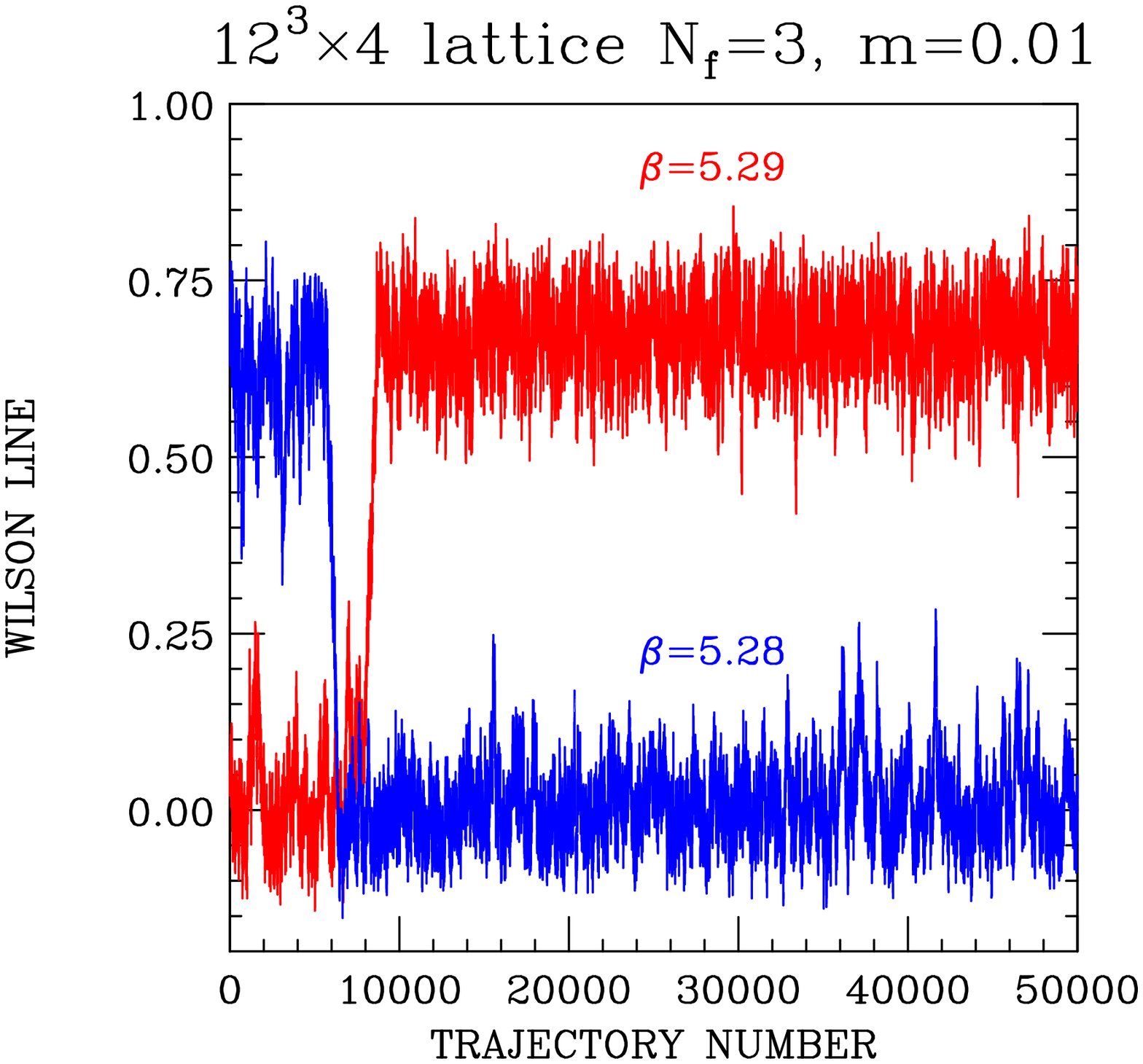}
\epsfxsize=3.2in
\epsffile{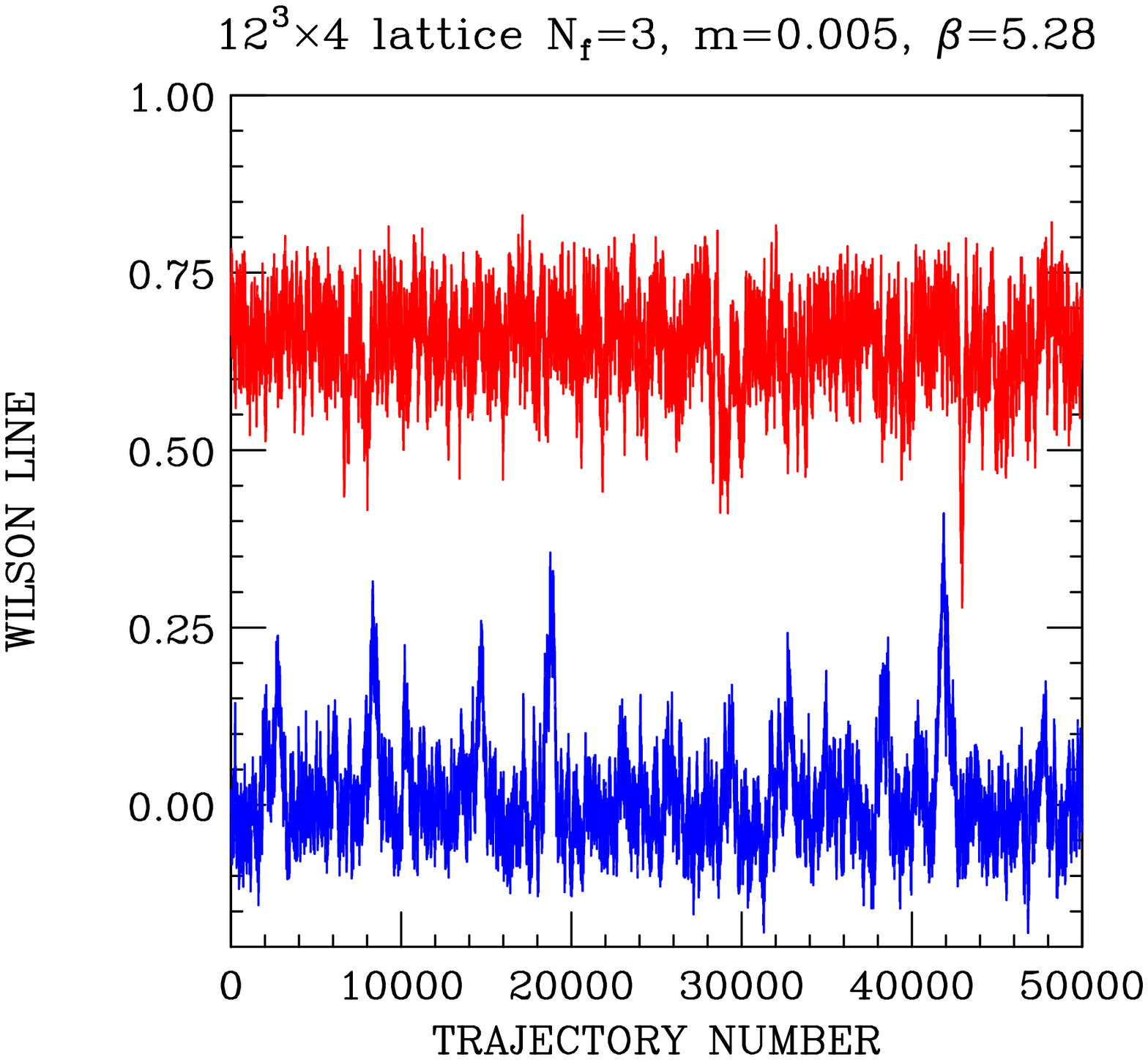}
\caption{a) Evolution of Wilson Line from a cold start at $\beta=5.3$ and from
a hot start at $\beta=5.29$ for $m=0.02$. b) Evolution of Wilson Line from a 
cold start at $\beta=5.29$ and from a hot start at $\beta=5.28$ for $m=0.01$.
c) Evolution of Wilson Line from cold and hot starts at $\beta=5.28$ for 
$m=0.005$.}
\label{fig:time4}
\end{figure}

Now we turn to more quantitative estimates of the position of the deconfinement
transitions for each of the 3 masses. We do this by examining the evolution of
the Wilson Lines with molecular-dynamic `time' near the transition and the
observation of tunneling. Such `time' histories are shown in 
figure~\ref{fig:time4}. Part (a) of this figure shows time evolutions for 
$m=0.02$. Starting from a cold start at $\beta=5.3$ we observe a tunneling from
a cold (small Wilson Loop) state to a hot (large Wilson Loop) state, after
which the system remains in this hot state for the rest of the run. Starting
from a hot state at $\beta=5.29$ the system tunnels to a cold state and
remains there. From this we conclude that $\beta=5.29$ is on the cold side of
the transition and $\beta=5.3$ is on the hot side of the transition. Our best
estimate of the position of the transition is thus $\beta=5.295(5)$. Similarly
from part (b) of this figure we deduce that the transition for $m=0.01$ lies
between $\beta=5.28$ and $\beta=5.29$ giving our best estimate as 
$\beta=5.285(5)$. Finally in part (c) we show the time evolution from hot and
cold starts at $\beta=5.28$ for $m=0.005$. Here there are no tunnelings in
either case for the duration (50,000 trajectories) of each run, from which we
conclude that the transition is close to $\beta=5.28$ -- $\beta=5.280(5)$.
The behaviour shown in this figure (\ref{fig:time4}) strongly suggests a 
first-order phase transition. However, in the absence of any finite-size
analysis, this observation is not conclusive.

\begin{figure}[htb]
\epsfxsize=6.0in
\epsffile{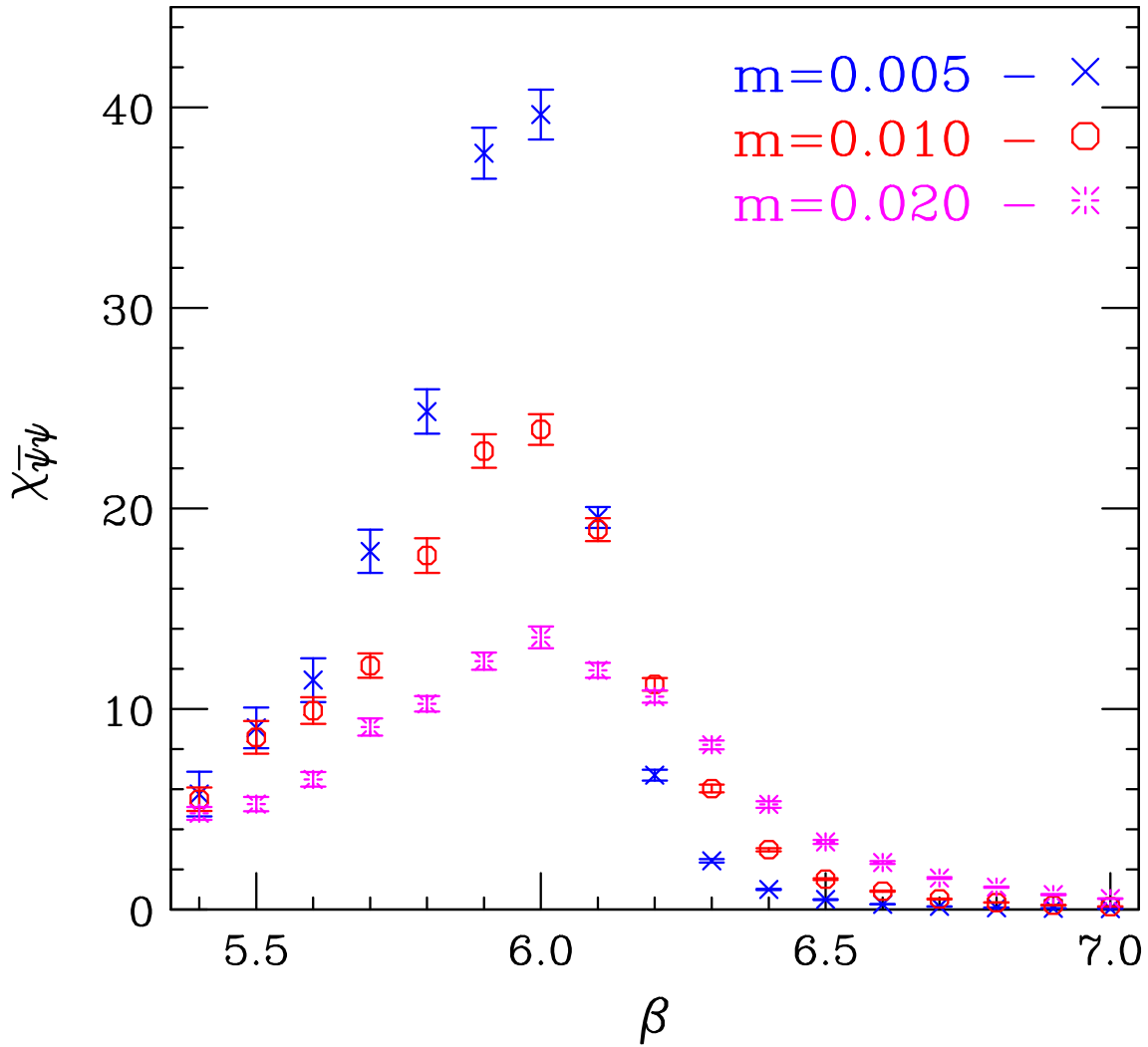}
\caption{Chiral susceptibilities $\chi_{\bar{\psi}\psi}$ as functions of $\beta$
on a $12^3 \times 4$ lattice for $m=0.02,0.01,0.005$, for a $\beta$ range which 
includes the chiral transition.}
\label{fig:chi4}
\end{figure}

The chiral condensate becomes small and shows a strong dependence on $m$ 
for large $\beta$. However, naive attempts to extrapolate to zero quark mass
depend strongly on the analytic form chosen for extrapolation. While we might
expect a first order transition for $N_f=3$, this is not obvious in the `data'.
We therefore examine the (disconnected part of the) chiral susceptibilities
discussed in section~2. These are plotted in figure~\ref{fig:chi4} for all
3 quark masses. $\chi_{\bar{\psi}\psi}$ shows a clear peak at $\beta=6.0$ for
each quark mass. The height of the peaks increases with decreasing quark mass,
as it should, since the susceptibility will diverge at the chiral phase
transition. Since the $\beta$ at the peak does not change with mass to the
resolution in $\beta$ of our simulations, we estimate that the chiral 
transition occurs at $\beta=\beta_\chi=6.0(1)$.

We also perform a second set of simulations which start from states with the
Wilson Line real and negative at $\beta=7.0$. We achieve this by starting
from a 2-flavour configuration with a real-negative Wilson Line also at 
$\beta=7.0$. Again we run for 10,000 trajectories at each $\beta$ until we
get close to the transition from the state with a negative Wilson Line to a
state with a complex Wilson Line with a phase close to $\pm 2\pi/3$, where we
increase this to 50,000 trajectories. For all 3 quark masses, this transition 
occurs at some $\beta$ in the range $5.5 < \beta < 5.6$. For $\beta$ below
$5.5$ we again simulate 10,000 trajectory runs. For $\beta \le 5.34$ at 
$m=0.02$ and $\beta \le 5.33$ at $m=0.01$ and $m=0.005$ we increase our run 
lengths to 50,000 trajectories. We perform closely spaced (in $\beta$) runs 
down to $\beta=5.3$. At $\beta=5.32$ and below for masses $m=0.02$, $m=0.01$, 
and $\beta=5.315$ and below for $m=0.005$ the complex Wilson Line state rapidly
decays to a positive Wilson Line state and remains there for the remainder of
the run. This we take as an indication that the states with complex Wilson lines
(and presumably those with negative Wilson Lines) are long-lived metastable
states. We also note that just above the $\beta$ values where this metastability
reveals itself, we observe tunnelings between the 2 complex-Wilson-Line states.

\section{$N_t=6$ simulations}

We simulate lattice QCD with 3 colour-sextet quarks on $12^3 \times 6$ lattices,
starting from states with a real positive Wilson Line at $\beta=7.0$, and from
states with a real negative Wilson Line at $\beta=7.0$. We use 3 different
quark masses $m=0.02$, $m=0.01$ and $m=0.005$ to enable us to access the
chiral ($m \rightarrow 0$) limit. The simulations at $\beta=7.0$ are started
from 2-flavour configurations with the same $\beta$ and the desired Wilson
Line orientation. Our runs for each $(\beta,m)$ for each initial orientation
of the Wilson Line are 10,000 trajectories in length away from the
transitions, increasing to 50,000 trajectories close to the deconfinement
transition, and close to the transitions from states with negative Wilson
Lines to states of complex Wilson Lines. We also found it necessary to 
increase our statistics at $m=0.01$ close to the chiral transition to 25,000
trajectories to accurately determine the position of the transition. In 
addition, we used 20,000 trajectories for each $(\beta,m)$ significantly below
the transition, where we only have runs connected to the start with a positive
Wilson line at $\beta=7.0$.

As $\beta$ is increased from its lowest value, chosen to be $\beta=5.3$ for 
each quark mass, the magnitude of the Wilson Line increases rapidly near
$\beta \approx 5.4$, signaling the deconfinement transition. Just above this
transition, there is a clear 3-state signal in the Wilson Line, where the
phase of this line is close to $0$ or $\pm 2\pi/3$. Hence we bin our `data'
according to which of these values the phase of the Wilson Line is closest.
We are then able to combine the 2 complex-Wilson-Line bins by conjugating
those Wilson Lines in the phase $-2\pi/3$ bin. Just above the deconfinement 
transition we observe tunnelings between the 3 states, with no indication that
any state is favoured. Hence, as far as these simulations are concerned, all
3 states appear stable (in the thermodynamic limit). In the region where this
3-state tunneling is observed, we combine the `data' from the 2 starts, giving
100,000 trajectories for each $(\beta,m)$.

\begin{figure}[htb]
\epsfxsize=6.0in
\epsffile{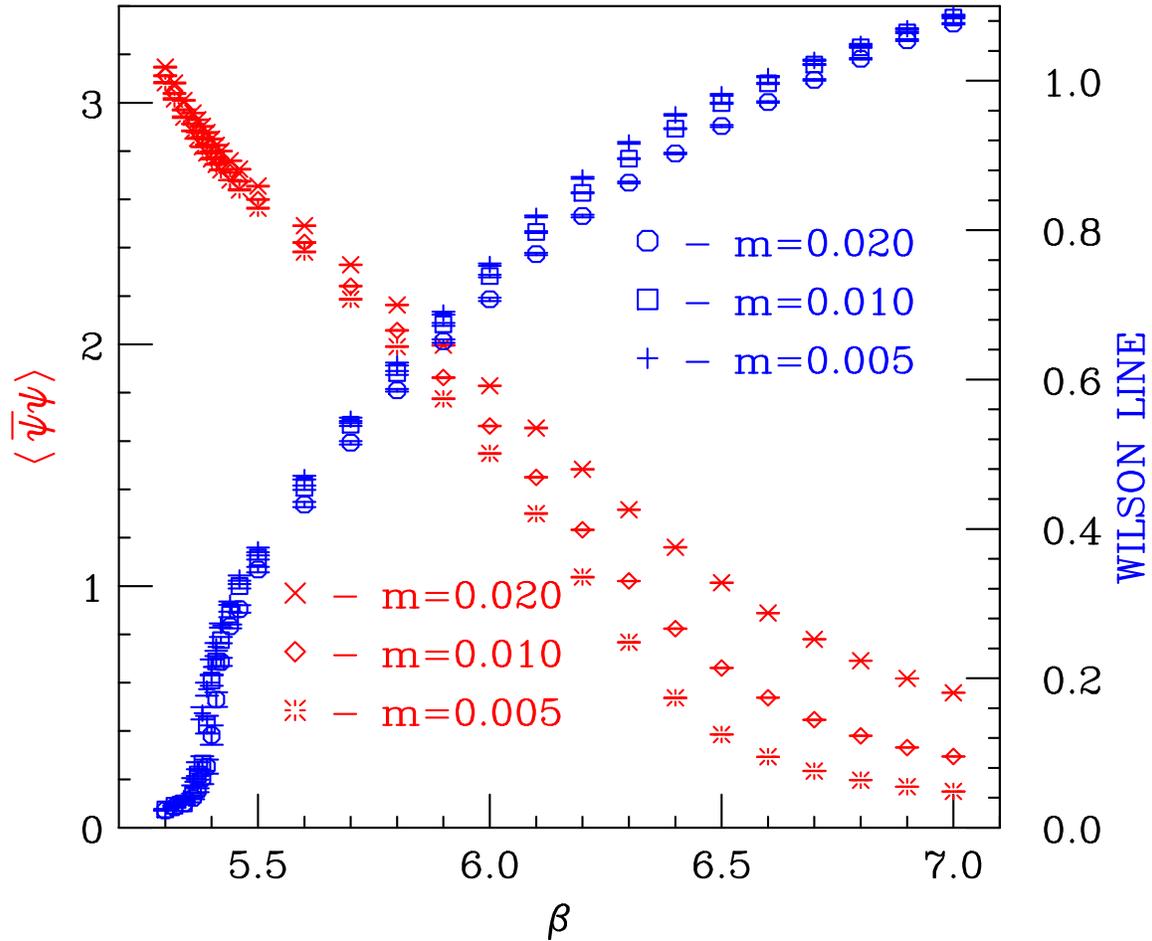}
\caption{Wilson Line and $\langle\bar{\psi}\psi\rangle$ as functions of 
$\beta=6/g^2$ on a $12^3 \times 6$ lattice, in the state with a real positive
Wilson Line, for $m=0.02,0.01,0.005$.}
\label{fig:rwil-psi6}
\end{figure}

Figure~\ref{fig:rwil-psi6} shows the Wilson Lines and chiral condensates
$\langle\bar{\psi}\psi\rangle$ for the state with a real positive Wilson Line,
for each of the 3 masses. The rapid increase in the Wilson Line near $\beta=5.4$
is clear. These Wilson Line values continue to increase over the whole range
of $\beta$s used in our simulations, and are expected to approach $3$ as
$\beta\rightarrow\infty$. The deconfinement transition has little effect on
the chiral condensates, which decrease over the range of $\beta$s under
consideration. At larger $\beta$ values these condensates become increasingly 
mass dependent, decreasing with decreasing mass, suggesting that they will 
vanish in the chiral limit. Just from looking at these graphs, one would
suspect that the chiral transition occurs at $\beta$ just below $6.5$, but it
is clear that a more reliable method, such as that provided by examining the 
chiral susceptibilities, is needed to determine this $\beta_\chi$ with any 
precision. This qualitative analysis does indicate, however, that the 
deconfinement and chiral transitions are still far apart.

\begin{figure}[htb]
\epsfxsize=6.0in
\epsffile{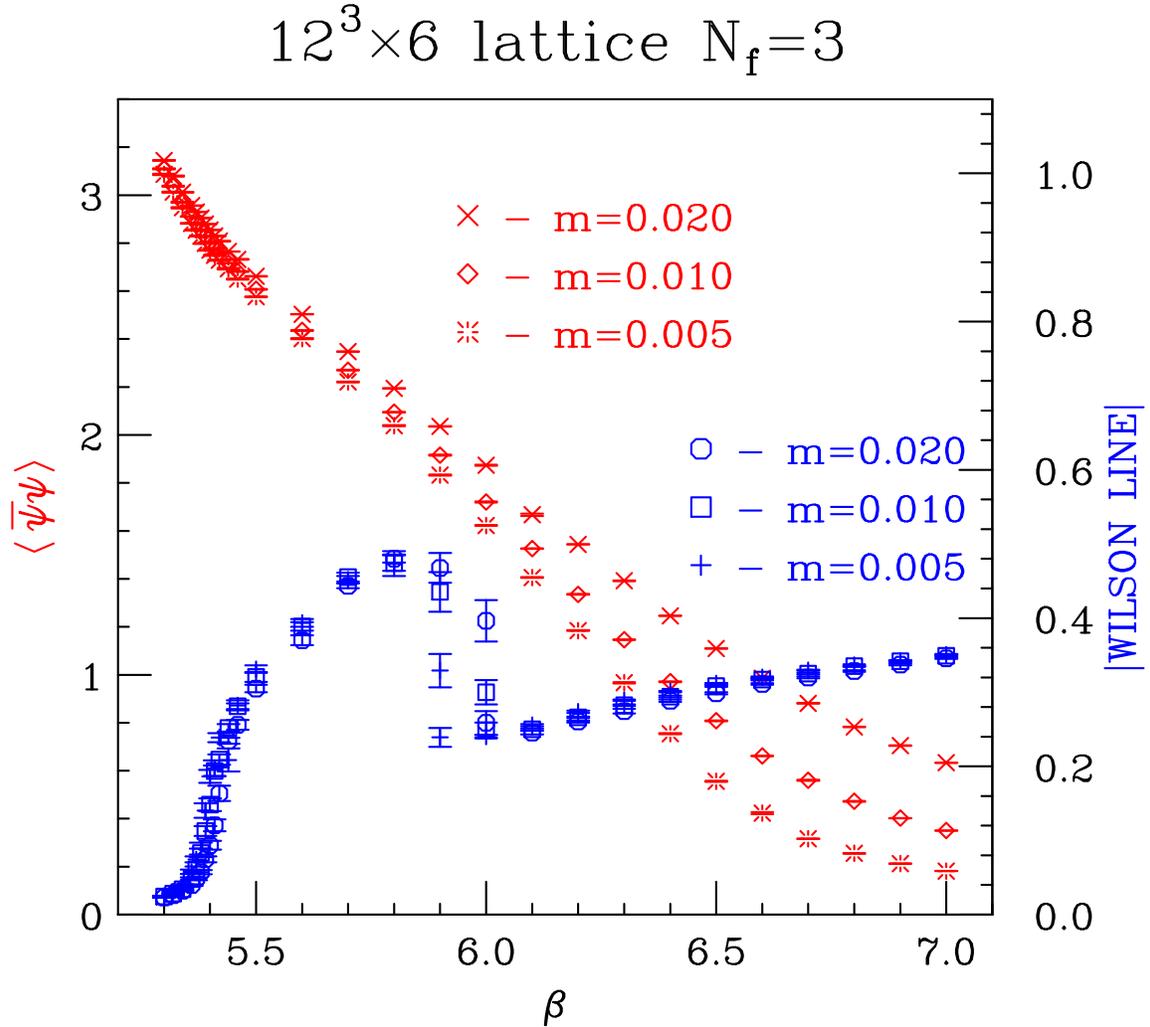}
\caption{Magnitude of the Wilson Line and $\langle\bar{\psi}\psi\rangle$ as 
functions of $\beta=6/g^2$ on a $12^3 \times 6$ lattice, in the states with
complex or negative Wilson Lines, for $m=0.02,0.01,0.005$.}\
\label{fig:cwil-psi6}
\end{figure}
 
We now turn our attention to the states with complex or real negative Wilson
Lines. First we note that states having Wilson Lines with phases $\pm 2\pi/3$
disorder to a state with a real negative Wilson Line for $\beta$ sufficiently
large. This occurs for $5.9 \lesssim \beta \lesssim 6.0$ for $m=0.005$ and at
slightly higher $\beta$s for the larger masses. In figure~\ref{fig:cwil-psi6}
we show the magnitudes of the Wilson Lines for states with complex or negative
Wilson Lines and the corresponding chiral condensates as functions of $\beta$
for each mass. Again the magnitude of the Wilson Line shows the deconfinement 
transition for $\beta \approx 5.4$. It also shows the transition from a complex
to a real negative Wilson Line for $\beta \approx 6$. The chiral condensate
barely responds to the deconfinement transition and shows indications that it
will vanish in the chiral limit for some $\beta$ in the neighbourhood of the
chiral transition for the positive Wilson Line state.

\begin{figure}[htb]
\epsfxsize=3.2in
\epsffile{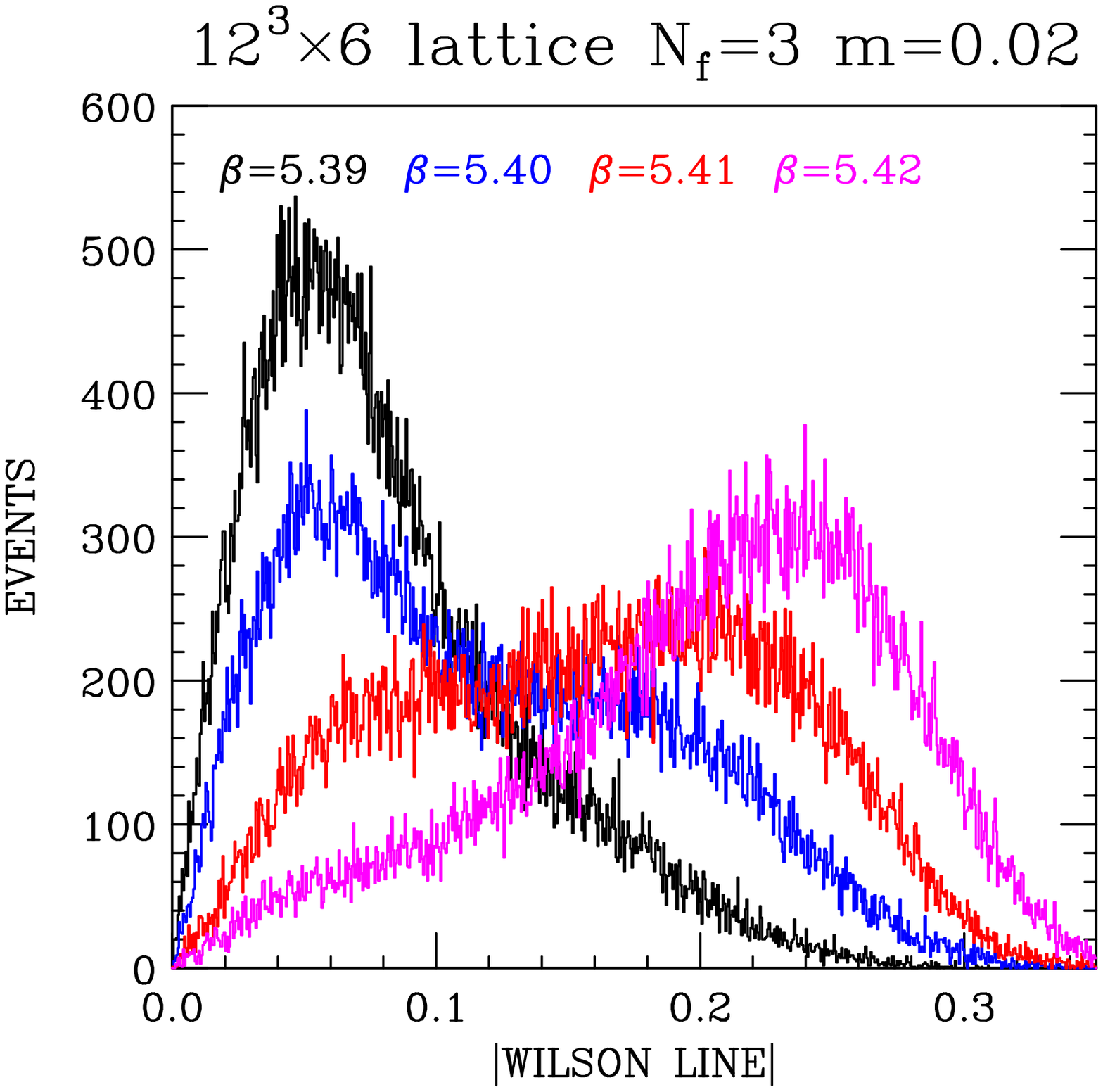}
\epsfxsize=3.2in
\epsffile{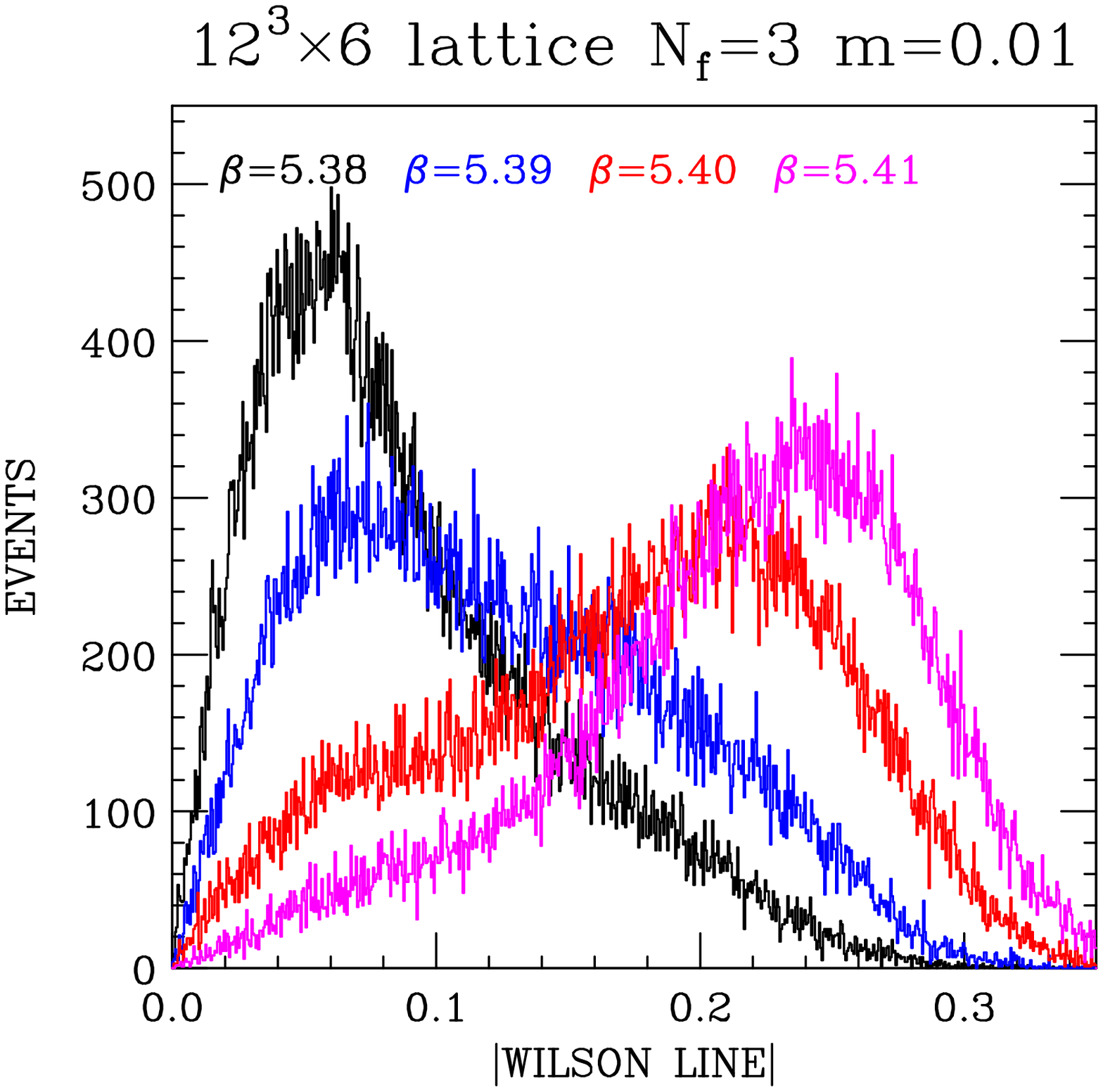}
\epsfxsize=3.2in
\epsffile{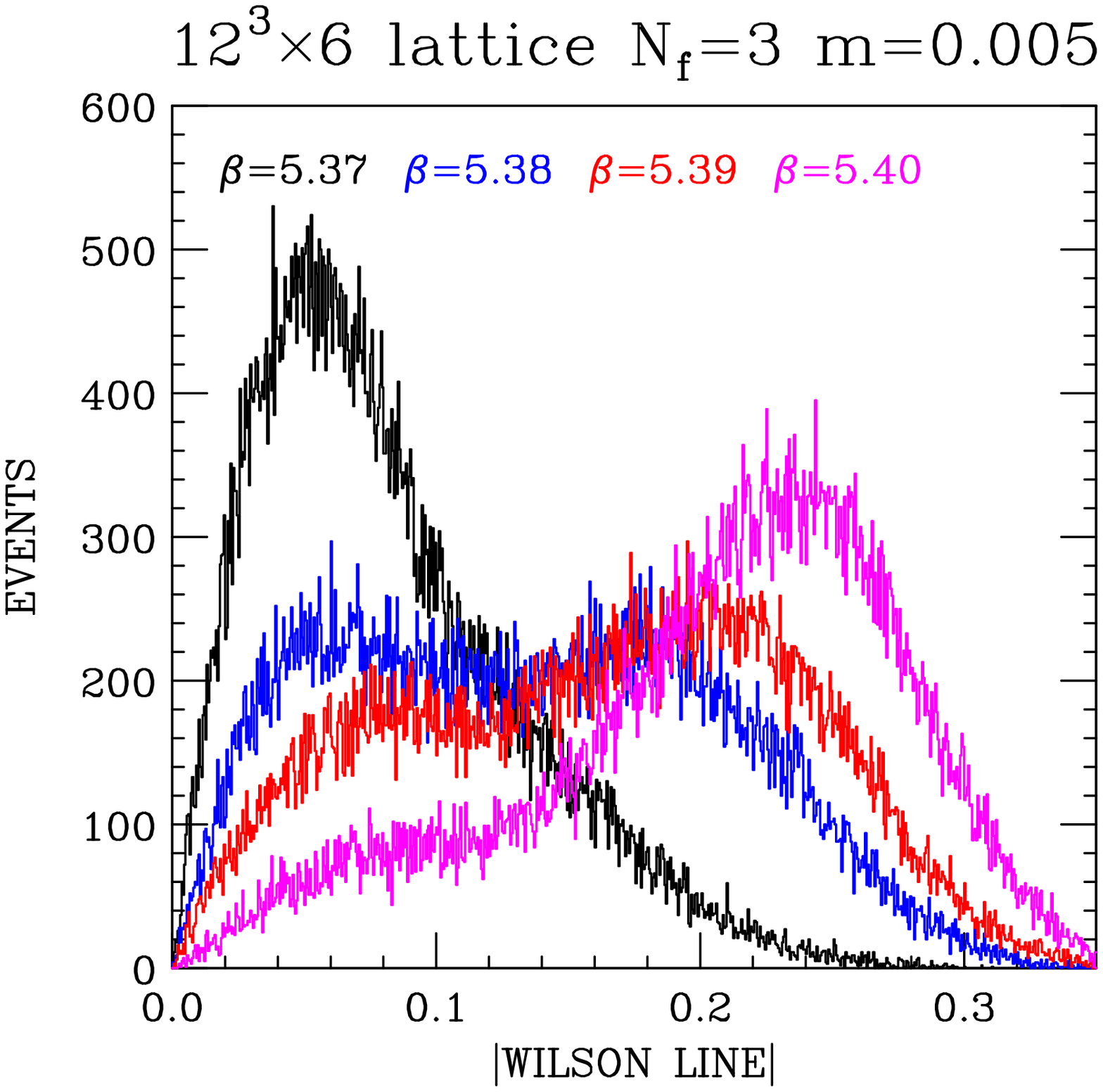}
\caption{Histograms of the magnitudes of Wilson Lines for $\beta$ values 
bracketing the deconfinement transition on a $12^3 \times 6$ lattice for 
a) $m=0.02$, b) $m=0.01$, c) $m=0.005$.}
\label{fig:wilhist}
\end{figure}

To more accurately estimate the positions of the deconfinement transition,
we histogram the magnitudes of the Wilson Lines in the vicinity of this
transition. We display such histograms in figure~\ref{fig:wilhist}. From these
we estimate that our transitions occur at $\beta=\beta_d$, where for $m=0.02$
$\beta_d=5.410(10)$, for $m=0.01$ $\beta_d=5.395(10)$ and for $m=0.005$
$\beta_d=5.385(10)$. If we assume  that below $\beta_d$ the magnitudes of the
real positive and complex Wilson Lines should be approximately the same, while
above they will differ, we get estimates of $\beta_d$ which are lower by 1--1.5
standard deviations.

\begin{figure}[htb]
\epsfxsize=6.0in
\epsffile{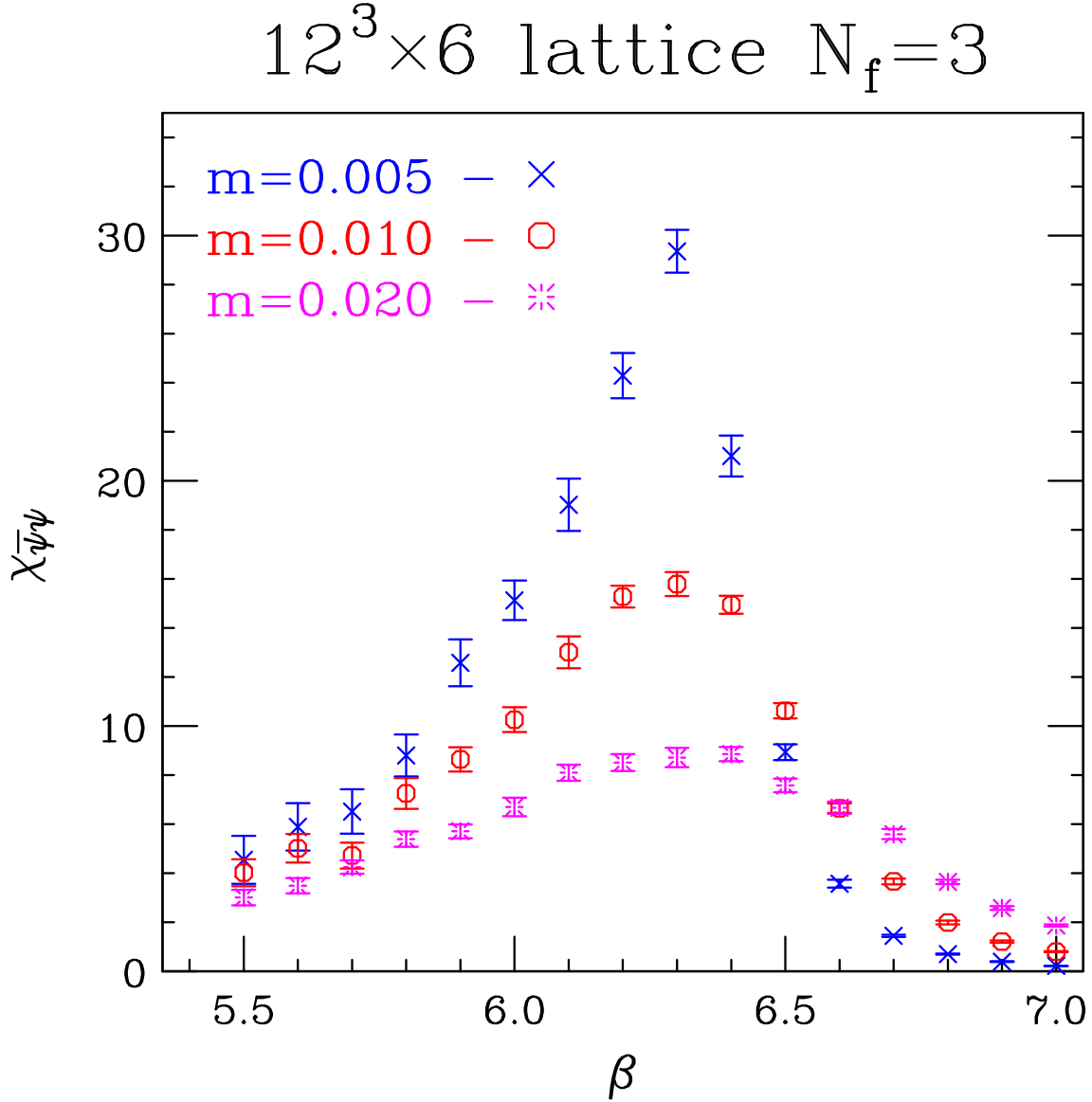}                                                       
\caption{Chiral susceptibilities $\chi_{\bar{\psi}\psi}$ as functions of $\beta$
on a $12^3 \times 6$ lattice for $m=0.02,0.01,0.005$, for a $\beta$ range 
which includes the chiral transition.}
\label{fig:chi6}                                                               
\end{figure}

As we have noted, determining the position of the chiral-symmetry restoration
phase transition directly from the chiral condensates themselves, is difficult
if not impossible, because we do not have a method of extrapolating the 
condensates to zero mass which we trust. Hence, to get an accurate estimate of
the $\beta$ value $\beta_\chi$ of the chiral phase transition, we turn to a 
consideration of the (disconnected) chiral susceptibility 
$\chi_{\bar{\psi}\psi}$. The position of the peak of $\chi_{\bar{\psi}\psi}$ as
a function of mass will approach $\beta_\chi$ as $m \rightarrow 0$. 
The value of $\chi_{\bar{\psi}\psi}$ at the peak will diverge (on a lattice of
infinite spatial volume) in this limit, making the position of the peak easier
to determine as the quark mass is decreased. Figure~\ref{fig:chi6} shows these
susceptibilities. What we observe is that the position of the peak does not
move very much as $m$ is decreased, so that the position of the peak at the
lowest mass $m=0.005$ should give a reasonable estimate of $\beta_\chi$. We
therefore estimate that $\beta_\chi=6.3(1)$. 

\section{Discussion and Conclusions}

We simulate the thermodynamics of QCD with 3 flavours of colour-sextet quarks 
on $12^3 \times 4$ and $12^3 \times 6$ lattices. Table~\ref{tab:trans} shows
our estimates for the positions of the deconfinement and chiral transitions.
\begin{table}[h]
%\parbox{0.25in}{$\:$}
%\parbox{2.5in}{
\centerline{
\begin{tabular}{|c|c|c|}
\hline
$N_t$          & $\beta_d$         & $\beta_\chi$             \\
\hline
4              &$\;$5.275(10)$\;$  &$\;$6.0(1)$\;$            \\
6              &$\;$5.375(10)$\;$  &$\;$6.3(1)$\;$            \\
\hline
\end{tabular}
}
\caption{$N_f=3$ deconfinement and chiral transitions for $N_t=4,6$. In each
case we have attempted an extrapolation to the chiral limit.}
\label{tab:trans}
\end{table}
First we note that the chiral and deconfinement transitions are far apart as
was the case for 2 flavours. Both transitions move to significantly larger
values of $\beta=6/g^2$ as $N_t$ is increased from $4$ to $6$. Since we expect
the 3-flavour theory to be conformal, the chiral transition should be a bulk
transition occurring at fixed $\beta$. This suggests that we are in the 
strong-coupling regime where the fermions play little part in the dynamics,
which is therefore quenched. We are thus seeing finite temperature transitions 
of the quenched theory. The move of both transitions towards weaker coupling is
controlled by the asymptotic freedom of the pure gauge theory. The bulk nature
of the continuum chiral transition will not reveal itself until $N_t$ is
sufficiently large for the chiral transition to emerge into the weak coupling
regime. The similarity in the behaviour of the $N_t=4$ and $6$ transitions
between the 2-flavour and 3-flavour theories led us to suspect that both
transitions might also lie in the strong-coupling regime for the 2-flavour
theory. For 2 flavours we have since obtained results of $N_t=8$ simulations
which, when combined with the $N_t=4$ and $6$ results support this conclusion
\cite{Kogut:2011ty}.
Thus 3-flavour simulations can help with the interpretation of the 2-flavour
simulations.

The phase structure of the 3-flavour theory is also similar to the 2-flavour
case. Above the deconfinement transition there is a 3-state signal, a remnant
of the $Z_3$ symmetry of the quenched theory. This is further evidence that
the theory is effectively quenched at these couplings. As in the 2-flavour case,
the 2 states with complex Wilson Lines disorder to a state with a real negative
Wilson Line for $\beta$ sufficiently large. For $N_t=4$ as for 2 flavours,
only the state with a positive Wilson Line appears stable; the other states
show signs of metastability. This contrasts with the $N_t=6$ case where all
states appear stable. The existence of such states where the Wilson Line has
phases $\pm 2\pi/3$ and $\pi$ in addition to that with phase $0$ has been
predicted by Machtey and Svetitsky and observed in their simulations with
Wilson fermions \cite{Machtey:2009wu}.

We are now extending these simulations to $N_t=8$, after which we will consider
extending them to $N_t=12$. This way we hope to observe the chiral transition
emerge into the weak-coupling domain. If the $N_f=2$ theory is QCD-like, we
hope that this should be adequate to distinguish the $N_f=2$ and $N_f=3$ 
theories. 

DeGrand Shamir and Svetitsky have been studying lattice QCD with 2 colour-sextet
quarks using improved Wilson quarks. The Lattice Higgs Collaboration has been
studying the 2 flavour theory using improved staggered quarks. Both these
collaborations have concentrated their efforts on the zero-temperature 
behaviour, except for some very early work.

\section*{Acknowledgements}

DKS is supported in part by the U.S. Department of Energy, Division of High
Energy Physics, Contract DE-AC02-06CH11357. 

This research used resources of the National Energy Research Scientific
Computing Center, which is supported by the Office of Science of the U.S.
Department of Energy under Contract No. DE-AC02-05CH11231. In particular,
these simulations were performed on the Cray XT4, Franklin and Cray XT5, Hopper,
both at NERSC. In addition this research used the Cray XT5, Kraken at NICS 
under XSEDE Project Number: TG-MCA99S015.

\end{document}